# Magnetotransport in PbTe nipi structures

J. Oswald, M. Pippan, G. Heigl, G. Span, T. Stellberger

Institut für Physik, Montanuniversität Leoben, Franz Josef Str. 18
A-8700 Leoben, Austria

**Abstract:** In this paper the 3D- and 2D- behaviour of wide single quantum wells which consist of one period of a PbTe nipi-structure is studied theoretically and experimentally. A simple model combines the 2D- subband levels and the 3D-Landau levels in order to calculate the density of states in a magnetic field perpendicular to the 2D plane. It is shown that at a channel width of about 500 nm on can expect to observe 3D- and 2D-behaviour at the same time. Finally the general design aspects for PbTe samples are discussed.

## 1. Introduction

PbTe is a narrow gap semiconductor with both the conduction and the valence band extrema located at the L points of the Brillouin zone [1,2]. The effective mass ellipsoids are oriented along the four <111> directions (Fig.1). The PbTe layer systems are grown on $BaF_2$ substrates by HWE (Hot Wall Epitaxy).

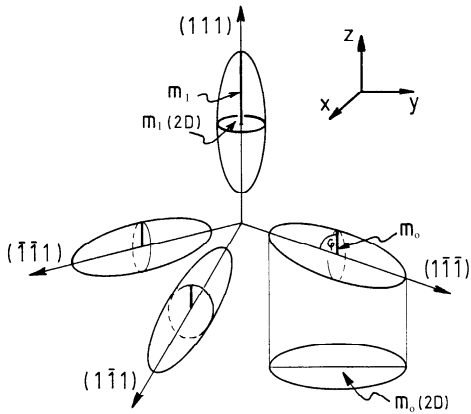

Fig.1. Effective mass ellipsoids of PbTe. With respect to the (111) growth direction there are two different effective masses for both the subband quantization: $m_l$, $m_o$ and the Landau level splitting: $m_l(2D)$, $m_o(2D)$.

The basic idea of doping superlattices is to create a periodic band-edge modulation in real space without changes in the chemical composition [3,4]. By using a typical doping level of $1 \times 10^{17}...3 \times 10^{17} cm^{-3}$ any potential modulation exceeding 10 meV leads to completely de-coupled potential wells in PbTe nipi-structures. The typical behaviour of such structures therefore can be investigated sufficiently by use of only a single period which in our case is a simple pnp-structure. In this pnp-structures the p-layers are designed to be non-depleted in order to screen the embedded n-channel from any influence from surface and interface states [5]. Selective contacts to the embedded n-channel are realised [6] in order to perform magneto transport experiments. The mobility in this selectively conducted structures is in the order of $3 \times 10^5 ... 1 \times 10^6$ cm$^2$V$^{-1}$s$^{-1}$ at a temperature of T = 2.2K. The extremely high dielectric constant ($\varepsilon$ = 500 - 1000) leads to wide quasi parabolic potential wells with a typical subband separation ranging from 1- 3 meV. A magnetic field perpendicular to the layers leads to a Landau level splitting which is large as compared to the subband splitting even at magnetic fields below B = 11 Tesla.

## 2. Model

For the modelling of the electronic structure in a magnetic field B||[111] it is a good approximation to assume the subband states and the Landau states to be separable. In this case the energy levels of the system can be simply written as:

$$E_{ij}^{l,o}(B) = E_i^{l,o} + E_j^{l,o}(B) \tag{1}$$

l denotes the longitudinal and o the oblique valleys, i denotes the subband level index and j the Landau level index.

The subband energies are calculated self consistently by a numerical solution of the Schrödinger equation within the effective mass approximation. The energy level system consists of all Landau levels (LL's) sitting on each of the subband levels. Since in wide quantum wells the Landau level separation is much higher than the subband splitting, a picture which considers the Landau levels to be split by the subband quantization is more convenient. Therefore simply the Landau levels resulting from a 3D-calculation according to [7] are used in Eqn. 1 (for more details on 3D-Landau levels of PbTe see also e.g. [8] or [9]). The modelling of the density of states (DOS) then is easily done by a superposition of individual Landau peaks with a broadening factor b. The broadening function in this approximation is assumed to be Gaussian. The total DOS as a function of energy E and magnetic field B then reads as follows:

$$DOS(E,B) = \frac{q_e \cdot B}{h} \sum_{i,j} \frac{1}{b \cdot \sqrt{\pi}} \cdot \left( e^{-\frac{(E - E_{ij}^l(B))^2}{b^2}} + 3 \cdot e^{-\frac{(E - E_{ij}^o(B))^2}{b^2}} \right) \tag{2}$$

The numerical factor 3 in the second term in brackets accounts for the threefold degeneracy of the oblique valleys. For a basic understanding one can consider two different idealized cases which will mix up in reality. One case is an ideal square potential and the second case is an ideal parabolic potential. If we have a square potential the subband energies depend on the square of the subband index. Therefore the lower subband levels are very close to each other and their associated Landau levels (of same Landau index) cause an overall enhancement of the density of states (DOS) near the subband ground state. The energy of the subband ground state is nearly identical to the position of the associated 3D - LL (LL of a three dimensional sample). At higher subband energies the subband levels are separated more and more and therefore a modulation of the DOS by the subband structure is more and more pronounced. In a parabolic potential well equally spaced subband levels are present. Therefore an ideal parabolic potential would not lead to an overall enhancement of the DOS at the position of the 3D - LL's. However, this idealized cases result from bare potentials and in real structures the situation will be different from both cases. The self consistent feedback of free electrons will change the potential as discussed in the following:

The effect of subband filling by adding electrons to an n-channel of a wide pnp-structure is limited by a Fermi energy which corresponds to the Fermi energy of bulk material at the same doping level. A further increase of the carrier sheet density therefore results in an increasing channel width and consequently a flat region in the middle of the electron channel is created. This behaviour starts in PbTe pnp structures which have a doping level of about $2\times10^{17}$cm$^{-3}$ at carrier sheet densities of about $6\times10^{12}$cm$^{-2}$. In this regime the subsequent filling of subbands by increasing the sheet carrier density is realised by a lowering of the subband energies while keeping the Fermi energy constant. The decrease of the subband energies happens by the increase of the channel width. In this way the subband structure turns more and more into the subband scheme of a square potential well. Consequently the lower part of the subband ladder looks like that of a square potential while the higher part will be dominated by the parabolic branches of the potential wells. For magneto transport experiments this means that there is an overall enhancement of the DOS at the energy position of the 3D-LL's and therefore also bulk like Shubnikov - de Haas (SdH) oscillations of the magneto resistance can be expected. At higher subband energies, which means the energy range between the 3D-LL's, a modulation of the DOS due to the subband structure can survive.

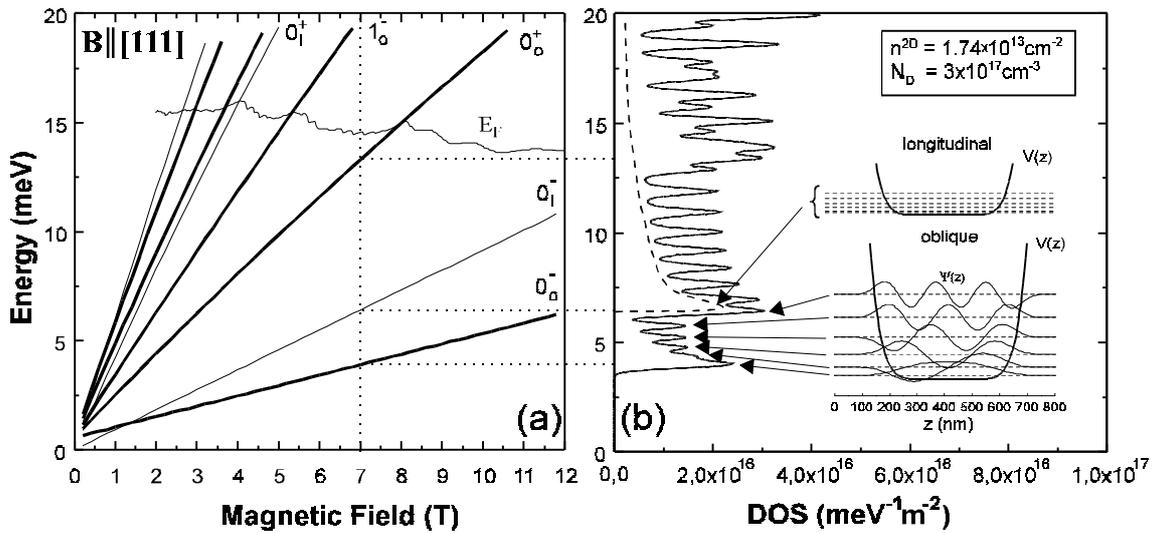

Fig 2. a) Landau levels of the subband ground states and Fermi energy versus magnetic field. b) density of states at B = 7 Tesla. Upper insert in b) potential and energy levels according to the longitudinal valley. Lower insert in b) potential, energy levels and wave functions according to the oblique valleys.

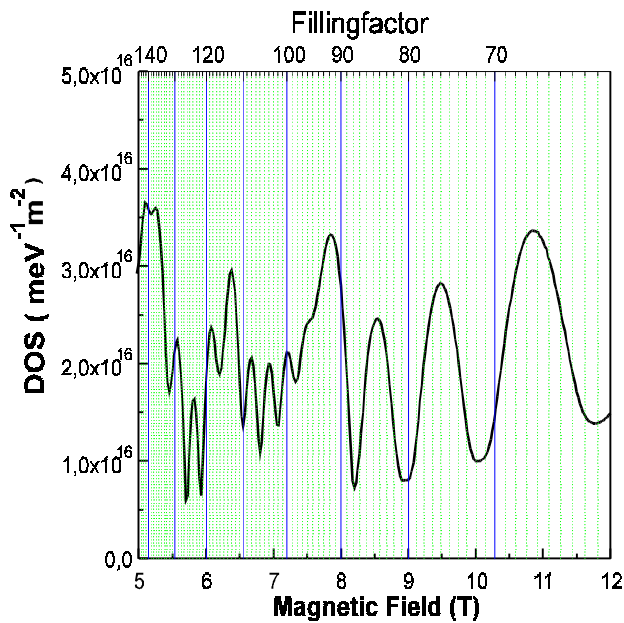

Fig.3: Density of states at the Fermi level versus magnetic field and filling factor according to Fig.2

Fig. 2 shows the result of a model calculation of a pnp-structure with parameters according to the sample used in the experimental section. For this sample the carrier sheet density is $n^{2D} = 1.74 \times 10^{13} \text{cm}^{-2}$ and the background doping level in the electron channel is $N_D = 3 \times 10^{17} \text{cm}^{-3}$. The resulting shape of the potential near the bottom in the electron channel can be seen from the insert in Fig.2b. In Fig.2a the Fermi energy versus magnetic field is strongly influenced by the Landau levels of the subband ground states. This behaviour is the same like in 3D samples. The reason can be seen from Fig.2b. In this figure the DOS is plotted for a constant magnetic field of B = 7 Tesla. An overall enhancement of the DOS near the position of the 3D-LL's levels can be understood from the overlap of the DOS- peaks of the lowest Landau-subband states. The LL's of the subband ground states at B = 7 Tesla are marked by the dotted lines guiding from Fig.2a to Fig.2b. Because of the small subband separation in the longitudinal valley their contribution to the DOS is more or less a uniform background which is indicated schematically by the dashed line in Fig.2b.

# 3. Experiments

The sample parameters in this experiment are those used for the model calculation in the pervious section. The shape of the sample is a conventional Hall bar geometry. The experiments where performed in an 11 Tesla super conducting magnet at pumped He[4] . Using a conventional lock-in technique, 3D like SdH oscillations dominate the magneto transport experiments (Fig.4a) without any additional structures in the data. In order to enhance the sensitivity of the experiment the set-up was extended to a double modulation arrangement:

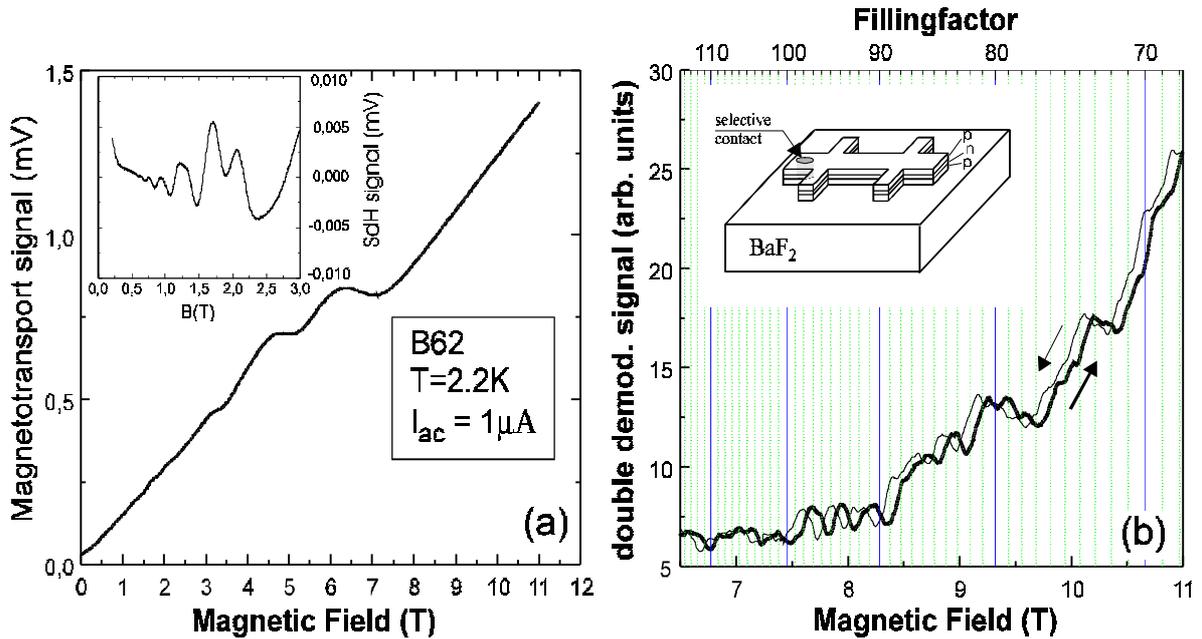

Fig.4 a) Magneto resistance data. Insert: magnification of the lower field range after subtraction of a linear slope. b) signal resulting from double modulation technique. Insert: Scheme of the sample configuration with one representative selective contact.

An additional constant ac-current at a different (lower) frequency is superimposed to the ac-current used for the normal magnetotransport measurement. In selectively contacted pnp-structures this additional current causes also an additional (small) voltage drop which occurs only in the n-layer. This voltage drop has the same effect as a gate voltage applied by isolated metal gates which cannot be used for our samples. The voltage drop within the n-channel causes a change of the pnp-potential and therefore causes also a change of the electron channel width. A modulation of the channel width, in turn, results in a modulation of the subband energies. The signal resulting from the conventional lock-in technique is used as the input signal for a second lock-in amplifier which is tuned to the modulation current. In this way the experiment is getting sensitive to structures in the DOS which are associated with the subband spectrum. As shown in Fig.4a a structure in the experimental data is found in the upper magnetic field range where no structure is observed in the direct signal. The signal is well reproducible by an up- and down- sweep of the magnetic field.

## 5. Discussion and conclusive remarks

The modulation of the DOS at the Fermi level (Fig.3) results from changes of the individual filling factors of the different involved subband systems. The carrier distribution between the different subband systems changes with the magnetic field. Since only the total filling factor is periodic in 1/B a plot of the data over 1/B cannot be used to check the periodicity. It is not possible to expect a particular period length to exist on the entire field range. In the investigated magnetic field range the model calculation shows that two ore three subband schemes overlap and the filling factors in that range are extremely high. Such a system is, of course, extremely sensitive to inhomogeneities or other influences and therefore it is not realistic to expect a one to one coincidence of the structures in the experimental and the simulated data. But the typical structural length scales can be found to be in agreement. A typical period length of $\Delta v \geq 3$ indicates that the oblique valleys dominate the subband structure. The model calculation includes also a strain induced energy shift between the longitudinal and the oblique valleys in the order of 1 meV [9]. Due to the sample parameters, in particular the high Fermi energy, this parameter is not very critical in this case. Concerning the general design aspects of PbTe pnp-samples one can say the following:

In order to get gaps in the DOS between the LL's all carriers have to be transferred to the oblique valleys. This is in principle always possible at sufficiently high magnetic fields (see LL's in Fig. 2). However, the required magnetic fields increase with increasing Fermi energy, which in turn depends on the doping level. Using a low Fermi energy the required magnetic field would shift to lower values (e.g. $B_{min} = 6$ Tesla at $N_D = 1 \times 10^{17} cm^{-3}$). But the disadvantage of a low Fermi energy is that if the quantum limit is reached (only lowest 3D-LL of the oblique valleys is occupied), the Fermi level enters the oblique subband scheme also at low subband energies and the subband separation might be not yet high enough to form gaps in the DOS (see Fig.2 b). From this point of view it is not good to aim at low filling factors and/or low carrier densities. The optimum situation is reached if the sample enters the quantum limit in the upper most part of the available magnetic field range so that the Fermi level is as far as possible above the subband ground state. That means that also the filling factor is kept at high numbers. This is opposite to 2D systems in hetero structures where one aims at low filling factors. But the reason for the different behaviour is simple: In ideal 2D systems made by heterostructures the subband splitting is much higher than the LL-splitting in the entire field range. Therefore the filling factor indicates the number of occupied LL's within the lowest subband. Going to low filling factors with increasing magnetic fields opens the gaps in the DOS more and more. In wide PbTe pnp-structures the gaps in the DOS are determined by the subband spectrum, which does not change much with the magnetic field. In this case the gaps between subband states are larger for higher subband indices ( = filling factors) provided that the sample has entered the quantum limit according to it's 3D-properties. The presented sample would enter this regime at magnetic fields between 15 and 20 Tesla.